\documentclass[fleqn,usenatbib]{mnras}

\usepackage{newtxtext,newtxmath}

\usepackage[T1]{fontenc}

\usepackage{CJKutf8}

\DeclareRobustCommand{\VAN}[3]{#2}
\let\VANthebibliography\thebibliography
\def\thebibliography{\DeclareRobustCommand{\VAN}[3]{##3}\VANthebibliography}

\usepackage{graphicx}	
\usepackage{amsmath}	
\usepackage{url}
\usepackage{comment}    
\usepackage{bm}
\usepackage{color}
\usepackage{multicol} 

\newif \ifja

\jafalse

\title[GR radiative transfer code: \texttt{CARTOON}]{3D Photon Conserving Code for Time-dependent General Relativistic Radiative Transfer\ :\ \texttt{CARTOON}}

\author[Mikiya M. Takahashi et al.]{
Mikiya M. Takahashi,$^{1}$\thanks{E-mail: mikiya@ccs.tsukuba.ac.jp}
Ken Ohsuga,$^{1}$
Rohta Takahashi,$^{2}$
Takumi Ogawa,$^{1}$
Masayuki Umemura,$^{1}$
\newauthor and Yuta Asahina$^{1}$
\\
$^{1}$Center for Computational Sciences, University of Tsukuba, 1-1-1 Tennodai, Tsukuba, Ibaraki 305-8577, Japan\\
$^{2}$National Institute of Technology, Tomakomai College, 443 Nishikioka, Tomakomai, Hokkaido 059-1275, Japan
}

\date{Accepted XXX. Received YYY; in original form ZZZ}

\pubyear{2022}

\begin{document}
\begin{CJK}{UTF8}{ipxm}
\label{firstpage}
\pagerange{\pageref{firstpage}--\pageref{lastpage}}
\maketitle

\begin{abstract}
We develop the 3-dimensional general relativistic radiative transfer code: \texttt{CARTOON} (Calculation code of Authentic Radiative Transfer based On phOton Number conservation in curved space-time) which is improved from the 2-dimensional code: \texttt{ARTIST} developed by Takahashi \& Umemura (2017).
In \texttt{CARTOON}, the frequency-integrated general relativistic radiative transfer equation is solved in a photon number-conserving manner,
and the isotropic and coherent scattering in the zero angular momentum observers (ZAMO) frame and the fluid rest frame is incorporated.
By calculating the average energy of photons, energy conservation of the radiation is also guaranteed.
With the test calculations in 2-dimensional and 3-dimensional space, we have demonstrated that the wavefront propagation in black hole space-time can be correctly solved in \texttt{CARTOON} conserving photon numbers. 
The position of the wavefront coincides with the analytical solution and the number of photons remains constant until the wavefront reaches the event horizon.
We also solve the radiative transfer equation on the geodesic reaching the observer's screen.
The time variation of the intensity map on the observer's screen can be simultaneously and consistently calculated with the time variation of the radiation field around the black hole.
In addition, the black hole shadow can be reproduced in moderately optically thin situations. 
\end{abstract}







\section{Introduction}
Radiation plays an important role in many astrophysical phenomena. 
One of the representative examples is a black hole (BH) accretion disc. 
In the BH accretion discs, the gravitational energy of the accreting matter is converted to thermal energy via the viscosity, and a part of that energy is released as radiation from the accretion discs. 
When radiation cooling is effective, geometrically thin discs are formed by decreasing the gas pressure \ \citep{Shakura1973}. 
In contrast, when radiation cooling is not effective, the BH accretion discs become of high temperature and geometrically thick \ \citep{Ichimaru1977, Narayan1994}.
In the case of an extremely high accretion rate, the geometrically thick discs are formed via the strong radiation pressure. The disc winds and jets are sometimes accelerated by the radiation force \ \citep{Begelman1983, Icke1989, Proga2000, Ohsuga2005, Takeuchi2010, McKinney2014, Takahashi2015, Kobayashi2018}.

In order to investigate the accretion discs, disc winds, and radiatively-driven jets, radiation hydrodynamics (RHD) simulations have been performed.
In the RHD simulations by \citet{Ohsuga2005}, the flux-limited diffusion (FLD) approximation is used. 
In the FLD method, the only zeroth moment equation is solved to update the radiation energy-momentum tensor. 
The radiation flux and the radiation stress are given as a function of the gradient of the radiation energy density. 
Recently, the M1 closure method is often employed in many RHD simulations and radiation magnetohydrodynamics (RMHD) simulations.
In the M1 closure method, the zeroth and first moment equations are solved to update the radiation energy-momentum tensor.
It is pointed out that both methods solve the radiation fields approximately, but the correct solutions are not always obtained in the optically thin regions or in the regions that the radiation field is extremely anisotropic \ \citep{Jiang2014a, Ohsuga2016}.
Although the general relativistic radiation megnetohydrodynamics (GRRMHD) simulations using M1 closure method have been performed \ \citep{Sadowski2013,McKinney2014,Takahashi2016}, correct solutions might not be obtained since the radiative transfer equation is not solved \ \citep[][]{Asahina2020, Asahina2022}.

The numerical method for the RMHD simulations, in which the time-dependent radiative transfer equation is directly solved, is suggested by \citet{Jiang2014a}.
Later, a fully special relativistic code is developed by \citet{Ohsuga2016}.
Recently, the RMHD simulation code in which the general relativistic radiative transfer (GRRT) equation is directly solved has been developed by \cite{Asahina2020}. 
Although they employ a grid-based method, solving the radiative transfer equation along the geodesic should further improve the accuracy.
\citet[][here after TU17]{Takahashi2017} has proposed a code called \texttt{ARTIST} that accurately solves for the wavefront propagation in the BH space-time by solving the time-dependent GRRT equation along the geodesics.
However, conservation of the photon number is not guaranteed in TU17. 
In addition, TU17 concentrated on radiation propagation on the equatorial plane so the radiation propagation in a 3-dimensional space cannot be solved in TU17.

In this paper, we develop a novel time-dependent GRRT calculation code: \texttt{\texttt{CARTOON}} improving the code developed by TU17. 
In \texttt{CARTOON}, the frequency-integrated GRRT calculations are performed with guaranteeing the conservation of the photon number. 
The numerical diffusion is very small compared to the grid-based GRRT solver \ \citep{Ogawa2021,Akaho2021}. 
The radiation energy-momentum tensor is evaluated based on the specific intensity. Since the average energy of photons is calculated, the radiation energy is conserved. 
We also compute the radiation transfer along the geodesic reaching the observer's screen. Thus, the time variation of the observed intensity map can be obtained simultaneously with the evolution of the radiation fields around a BH. 
In Section \ref{sec:eq}, basic equations are explained. 
In Section \ref{sec:sec2}, the numerical method is explained. 
In Section \ref{sec:sec3}, we show the results of the test calculations. Finally, the results are summarized in Section \ref{sec:sec4}.

Throughout the present study, we use the units $c=G=h=1$. 
Greek indices $\alpha$, $\beta$, $\gamma$, and $\delta$ denote $0-3$, and Latin ones $l$ and $m$ denote $1-3$. 
The hat ( $\hat{\ }$ ) and tilde ( $\tilde{\ }$ ) denote the physical quantities measured in the zero angular momentum observers (ZAMO) frame and the fluid rest frame.

\section{Basic Equation}
\label{sec:eq}
\subsection{Geodesic equation}
In the BH space-time, the geodesic equations are as follows:  
\begin{align} 
    &\frac{dx^\alpha}{d\lambda} = p^\alpha, \label{geoeq1} \\
    &\frac{d p_\alpha}{d\lambda} = -\frac{1}{2} \frac{\partial g^{\beta \gamma}}{\partial x^\alpha}p_\beta p_\gamma, \label{geoeq2}
\end{align}
where $x^\alpha$, $\ p^\alpha (p_\alpha)$, $\lambda$, and $g_{\alpha \beta} (g^{\alpha \beta})$ are the coordinates of the photon, the four momentum of the photon, the affine parameter, and the metric tensor. 
The geodesic equations (\ref{geoeq1}) and (\ref{geoeq2}) are equivalent to the standard form given in \citet{Gravitation} and used in \citet{Carter1963, Broderick2003, Takahashi2017}.
The 3+1 formulation of the metric in the spherical Kerr--Schild coordinate ($x^0=t$, $x^1=r$, $x^2=\theta$, $x^3=\phi$) is given by
\begin{equation} \label{metric1}
    ds^2 = g_{\alpha \beta} dx^\alpha dx^\beta = -\alpha^2 dt^2 + \gamma_{lm} (dx^l + \beta^l dt) (dx^m + \beta^m dt),
\end{equation}
where $ds$ is the line element of the BH space-time. The non-zero components of the lapse function $\alpha$, the shift vector $\beta^l$, and the spatial metric $\gamma_{lm}$ are
\begin{align} \label{metric2}
    &\alpha = \left( 1+\frac{2Mr}{\Sigma} \right)^{-\frac{1}{2}},\ \beta^r = \frac{2Mr}{\Sigma + 2Mr}, \nonumber \\
    &\gamma_{rr} = 1+\frac{2Mr}{\Sigma},\ \gamma_{\theta \theta} = \Sigma,\ \gamma_{\phi \phi} = \frac{A \sin^2 \theta}{\Sigma},\nonumber \\
    &\gamma_{r \phi} = \gamma_{\phi r} = -a\sin^2 \theta \left( 1+\frac{2Mr}{\Sigma} \right),
\end{align}
where
\begin{align} \label{metric3}
&\Sigma = r^2+a^2\cos^2\theta, \nonumber \\
&A=\left( r^2+a^2 \right)^2-a^2\Delta \sin^2 \theta,\nonumber \\
&\Delta = r^2-2Mr+a^2.
\end{align}
$M$ is the BH mass and $a$ is the dimensionless BH spin parameter. 
The tetrad vectors to transform coordinates between the global Kerr--Schild frame and the ZAMO frame are as follows \ \citep{Takahashi2008}:
\begin{align} \label{tetrad}
    &e^{\hat{t}}_{\alpha}      = \left( \alpha, 0, 0, 0 \right),\  e^{\hat{r}}_{\alpha}      = \frac{1}{\sqrt{\gamma_{rr}}} \left( \beta_r,\gamma_{rr},0,\gamma_{r\phi} \right), \nonumber \\
    &e^{\hat{\theta}}_{\alpha} = \left( 0, 0, \sqrt{\gamma_{\theta \theta}}, 0 \right),\ 
    e^{\hat{\phi}}_{\alpha}   = \left( 0, 0, 0, \frac{1}{\sqrt{\gamma^{\phi \phi}}} \right), \\
    &e^{\alpha}_{\hat{t}}      = \frac{1}{\alpha} \left( 1,-\beta^r,0,0 \right),\  
    e^{\alpha}_{\hat{r}}      = \left( 0, \frac{1}{\sqrt{\gamma_{rr}}}, 0, 0 \right), \nonumber \\
    &e^{\alpha}_{\hat{\theta}} = \left( 0, 0, \frac{1}{\sqrt{\gamma_{\theta \theta}}},0 \right),\ 
    e^{\alpha}_{\hat{\phi}}   = \sqrt{\gamma^{\phi \phi}} \left( 0, \frac{\gamma^{r\phi}}{\gamma^{\phi \phi}}, 0, 1 \right).
\end{align}

\subsection{General Relativistic Radiative Transfer Equation}
The GRRT equation along the geodesic is as follows:
\begin{equation} \label{GRRTeq}
    \frac{D \mathcal{I}}{d \lambda} = (\mathcal{E}_{\rm e} + \mathcal{E}_{\rm s}) - (\mathcal{A}_{\rm a} + \mathcal{A}_{\rm s}) \mathcal{I},
\end{equation}
where $\mathcal{I}$, $\mathcal{E}_{\rm e}$, $\mathcal{E}_{\rm s}$, $\mathcal{A}_{\rm a}$, and $\mathcal{A}_{\rm s}$
are the invariant distribution function of photons in a 6-dimensional phase space, the invariant emission coefficient, the invariant emission coefficient for scattering, the invariant absorption coefficient, and the invariant scattering coefficient respectively. 
The differentiation along the geodesic, $D/d\lambda$, is given by
\begin{equation} \label{Ddlambda}
\frac{D}{d\lambda} = \frac{\partial}{\partial \lambda} - \Gamma^\gamma_{\alpha \beta} p^\alpha p^\beta \frac{\partial}{\partial p^\gamma},
\end{equation}
with $\Gamma^\gamma_{\alpha \beta}$ being the Christoffel symbol. $\mathcal{I}$, $\mathcal{E}_{\rm e}$, $\mathcal{E}_{\rm s}$, $\mathcal{A}_{\rm a}$, and $\mathcal{A}_{\rm s}$ satisfy the following relations:
\begin{align} 
    &\mathcal{I} = \frac{I_\nu}{{\nu}^3}, \label{huhenI} \\
    &\mathcal{E}_{\rm e} = \frac{j_{\nu}}{{\nu}^2}, \label{huhenEe} \\
    &\mathcal{E}_{\rm s} = \int \int \sigma_{{\nu}^{\prime}} I_{{\nu}^{\prime}} \zeta(x^\alpha;p^{{\alpha}^{\prime}} \rightarrow p^{\alpha}) \frac{d{\nu}^{\prime}}{{{\nu}}^{\prime}} d\Omega^{\prime}, \label{huhenEs} \\
    &\mathcal{A}_{\rm a} = \nu \kappa_{\nu}, \label{huhenAa} \\
    &\mathcal{A}_{\rm s} = \nu \sigma_{\nu} \label{huhenAs},
\end{align}
where $I_{\nu}$, $j_{\nu}$, $\kappa_{\nu}$, and $\sigma_{\nu}$ are the intensity, the emission coefficient, the absorption coefficient, and the scattering coefficient at the frequency $\nu$ in the local Minkowski frame.
The solid angle and the scattering probability function in the local Minkowski frame are given by $\Omega$ and $\zeta$.
Here, the physical quantities before (after) scattering are shown with (without) prime.
In addition, the photon number $N$ is related to the invariant distribution function $\mathcal{I}$ as 
\begin{equation} \label{ItoN}
    N = \int \mathcal{I} d\mathcal{V}.
\end{equation}
Here, the invariant volume element in a 6-dimensional phase space $d\mathcal{V}$ is described as
\begin{equation} \label{dV}
    d\mathcal{V}={\nu}^2 d{\nu} d{\Omega} d{V},
\end{equation}
where $dV$ is the volume element of a 3-dimensional real space. 

\subsection{Radiation energy-momentum tensor}
The radiation energy-momentum tensor in the ZAMO frame $R^{\hat{\alpha} \hat{\beta}}$ is evaluated by
\begin{equation} \label{Rmunuhat}
    R^{\hat{\alpha} \hat{\beta}} = \int I_{\hat{\nu}} \frac{p^{\hat{\alpha}}}{p^{\hat{t}}} \frac{p^{\hat{\beta}}}{p^{\hat{t}}} d{\hat{\nu}} d{\hat{\Omega}},
\end{equation}
where $I_{\hat{\nu}}$, $\hat{\nu}$, and $d\hat{\Omega}$ are the intensity, the frequency, and the infinitesimal solid angle in the ZAMO frame. 
The radiation energy-momentum tensor in the global Kerr--Schild frame $R^{\alpha \beta}$ is evaluated by $R^{\hat{\alpha} \hat{\beta}}$ and the tetrad vectors $e^{\alpha}_{\hat{\beta}}$ as
\begin{equation} \label{Rmunu}
    R^{\alpha \beta} = e^{\alpha}_{\hat{\gamma}} e^{\beta}_{\hat{\delta}} R^{\hat{\gamma} \hat{\delta}}.
\end{equation}
Also, the radiation energy-momentum tensor in the fluid rest frame $R^{ \tilde{\alpha} \tilde{\beta}}$ is obtained by 
\begin{equation} \label{Rmunutilde}
    R^{\tilde{\alpha} \tilde{\beta}} = \Lambda^{\tilde{\alpha}}_{\hat{\gamma}} \Lambda^{\tilde{\beta}}_{\hat{\delta}} R^{\hat{\gamma} \hat{\delta}}.
\end{equation}
Here,
\begin{equation} \label{lorentz}
    \Lambda^{\tilde{\alpha}}_{\hat{\beta}} = 
    \left(  
      \begin{matrix}
          \gamma_{\rm v} & -\gamma_{\rm v} v^{\hat{l}} \\
          -\gamma_{\rm v} v^{\hat{l}} & \delta^l_m + (\gamma_{\rm v} -1) \frac{v^{\hat{l}} v^{\hat{m}}}{v^2}
      \end{matrix}
    \right)
\end{equation}
is the Lorentz transformation matrix, 
where $\delta^l_m$ is the Kronecker delta, $\gamma_{\rm v} = (1-v^2)^{-1/2}$ is the Lorentz factor, $v^{\hat{l}}$ is the three velocity of the fluid in the ZAMO frame and $v^2=v^{\hat{l}} v_{\hat{l}}$.
\section{Numerical method}
\label{sec:sec2}

We solve the geodesic equations (\ref{geoeq1}) and (\ref{geoeq2}), and set the geodesic grids for preparation.
We also divide the computational domain into the spatial cells which are independent of the geodesic grids.
We solve the GRRT equation by the photon number conserving scheme.
The radiation energy-momentum tensor is then calculated. 
Finally, we calculate the time variation of the intensity map on the distant observer's screen.

\subsection{Geodesic grid and spatial cell}
\label{sec:geogrid}
In this paper, we generate the geodesics in the whole of the computational domain defined in the range of $r_{\rm in} \leq r \leq r_{\rm out}$, $0 \leq \theta \leq \pi$, $0 \leq \phi \leq 2\pi$. Here, the radius of the inner and outer boundaries of the computational domain is set to be $r_{\rm in}=r_{\rm h}$ and $r_{\rm out}=25 r_{\rm g}$ where $r_{\rm g}=M$ and $r_{\rm h}=M+\sqrt{M^2-a^2}$ are the gravitational radius and the event horizon of the BH, respectively.

Here, we integrate the geodesic equations from start points on the inner and outer boundaries.
The coordinates of the start points are 
{\color{black} $(t,r,\theta,\phi)=(0,r_{\rm h} + \epsilon,\theta_p,\phi_q)$ and $(0,r_{\rm out},\theta_p,\phi_q)$ with $\epsilon$ being $10^{-5} r_{\rm g}$}.
We set the $\theta$ and $\phi$ coordinate of start points as $\theta_p = p\Delta \theta \ (1\leq p \leq N^{\rm geo}_{\theta}-1,\ p{\rm \ is\ an\ integer})$ and $\phi_q = q \Delta \phi \ (0\leq q \leq N^{\rm geo}_\phi-1,\ q{\rm \ is\ an\ integer})$, 
where $N^{\rm geo}_{\theta} = 24$ ($\Delta \theta = \pi/N^{\rm geo}_{\theta}$) and $N^{\rm geo}_\phi = 64 \ (128)$ for 3-dimensional (2-dimensional) calculations ($\Delta \phi = 2\pi/N^{\rm geo}_\phi$).
Therefore, the total number of start points is $2\times (N^{\rm geo}_\theta-1) \times N^{\rm geo}_{\phi}$.

To determine the four momentum of the photon at start points, we select the angles $\bar{\theta}$ and $\bar{\phi}$ in the ZAMO frame. 
Here, $\bar{\theta}$ indicates the angle between the coordinate axis in $r$ direction and the space components of the four momentum of the photon $p^{\hat{l}}$ in the ZAMO frame, and $\bar{\phi}$ is the angle between the coordinate axis in $\phi$ direction and $p^{\hat{l}}$ 
(see also Fig. 1 of \citet{Shibata2014}). 
The space components of the four momentum of the photon is 
obtained from $\bar{\theta}$ and $\bar{\phi}$ as
\begin{equation} \label{thbphb}
    p^{\hat{r}} = p^{\hat{t}} \cos \bar{\theta},\  p^{\hat{\theta}} = -p^{\hat{t}} \sin \bar{\theta} \cos \bar{\phi},\ 
    p^{\hat{\phi}} = -p^{\hat{t}} \sin \bar{\theta} \sin \bar{\phi},
\end{equation}
where the time component $p^{\hat{t}}$ is determined by $p_t$ which can take an arbitrary value because of the symmetry of the Kerr space-time. 
Then we obtain the four momentum of the photon in the global Kerr--Schild coordinate as $p^{\alpha} = e^{\alpha}_{\hat{\beta}} p^{\hat{\beta}}$. 

\begin{figure}
	\includegraphics[width=\columnwidth]{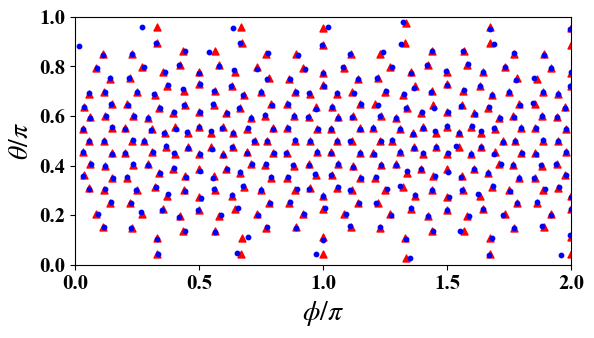}
    \caption{
    Blue dots show $\theta,\phi$ coordinate at the end points of scattering orbit from $(r,\theta,\phi)=(r_{\rm out}, \pi/2, 0)$. Red triangles show the centre of the curved triangles on the outer boundary. 
    }
    \label{fig:fig3}
\end{figure}

We set the geodesic points using the following procedures.
In order to efficiently prepare the geodesics for radiative transfer calculations, we equally divide the inner and outer boundaries of the computational domain into $N_{\rm d}(=320)$ curved triangles based on vertices of an icosahedron
\footnote{ {\color{black}  \texttt{Icosahedron-based Geodesic Dome}: \url{https://bitbucket.org/kohji/icosa} } }. 
We randomly choose one pair of $\bar{\theta},$ $\bar{\phi}$ and calculate a geodesic toward the inside of the computational domain.
Then, we numerically solve the geodesic equations (\ref{geoeq1}) and (\ref{geoeq2}) by the 8th order Runge-Kutta method until the geodesic reaches either the inner or outer boundary. 
We judge which curved triangle the intersection of the boundary and a geodesic (end point) belongs to. 
After that, we randomly select $\bar{\theta}$, $\bar{\phi}$ again, and calculate in the same way from the same start point. 
This procedure is repeated until the geodesics reach all curved triangles of the inner and outer boundaries.
If the multiple geodesics reach the same curved triangle, only the geodesic closest to the centre of the curved triangle is adopted.
In this way, $2N_{\rm d}$ geodesics starting from a start point are obtained
(here, we note that the geodesics that orbit more than three times around the BH in the $\phi$ direction are not adopted).
Blue dots in Fig. \ref{fig:fig3} indicate 
$\theta$ and $\phi$ coordinate at the end points of the geodesics from $(r,\theta,\phi)=(r_{\rm out}, \pi/2, 0)$. We find that the end points are located near the centres of all curved triangles (red triangles).

\begin{figure}
	\includegraphics[width=\columnwidth]{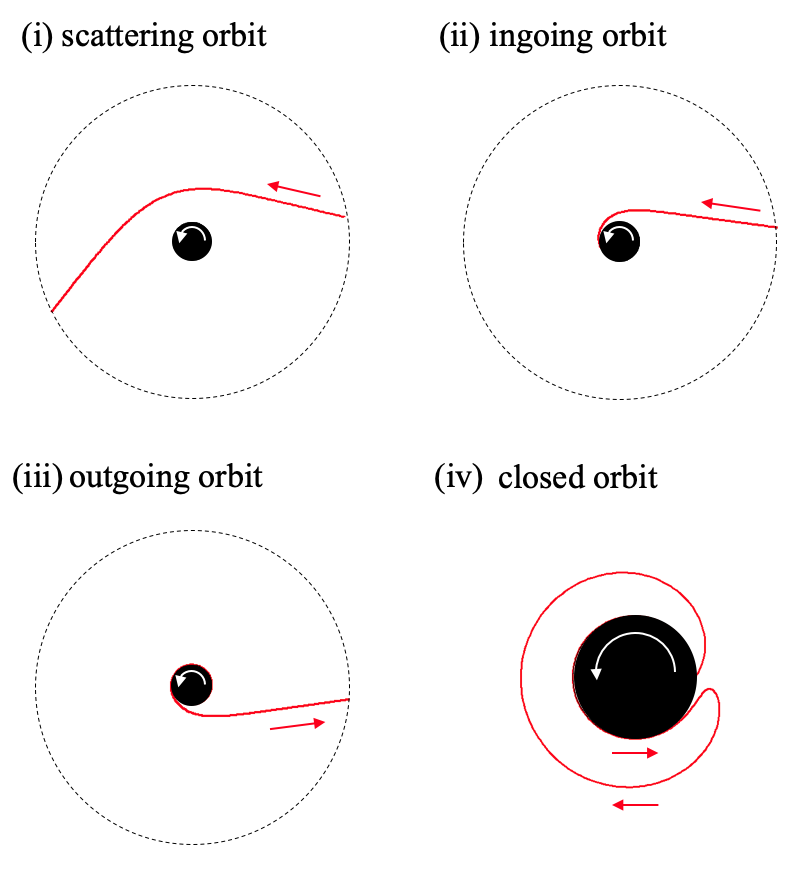}
    \caption{
    Patterns of the geodesics around BH: (i) scattering orbit, (ii) ingoing orbit, (iii) outgoing orbit, (iv) closed orbit. Black circle, white arrow, red solid line, red arrow, and black dotted line denote BH, the direction of BH spin, the geodesic, the direction of the photon propagation, and the outer boundary. 
    }
    \label{fig:fig2}
\end{figure}

We perform the above procedures for all start points and obtain  $2N_{\rm d} \times 2 (N^{\rm geo}_{\theta}-1) N^{\rm geo}_{\phi}$ geodesics in total.
{\color{black} The resulting geodesics are classified in 4 patterns as described} in Fig. \ref{fig:fig2}: (i) scattering orbits with both a start point and an end point on the outer boundary, (ii) ingoing orbits with a start point on the outer boundary and an end point on the inner boundary,  (iii) outgoing orbits with a start point on the inner boundary and an end point on the outer boundary, (iv) closed orbits with both a start point and an end point on the inner boundary. 
{\color{black}
Geodesics that orbit one or two times around the BH are already included in 4 patterns of geodesics described in Fig. \ref{fig:fig2}. However, we don't use the geodesics that orbit the BH more than three times. The accuracy of this method is discussed in Appendix \ref{sec:AppA}.
}

Then, points are set on the geodesics with equal intervals of $0.2M(=\Delta t = p^t \Delta \lambda)$. 
We call these points "geodesic grids" which are used by solving the GRRT equation numerically.
The geodesic grids are labeled $j=0,1,2,\cdots$ from a start point to an end point. 
Hereafter, the $j$-th point on the $i$-th geodesic is denoted as the geodesic grid $(i,j)$. Here, each geodesic is assigned a number $i$, but the method of assigning $i$ is arbitrary.
Next, we divide the computational domain into $N_r\times N_\theta \times N_\phi$ spacial cells. 
In this paper, subscripts $k_r$, $k_\theta$, $k_\phi$ denote the grid point of the spatial cell in the $r$, $\theta$, $\phi$ directions.

\begin{figure*}
  \begin{center}
	\includegraphics[width=\linewidth]{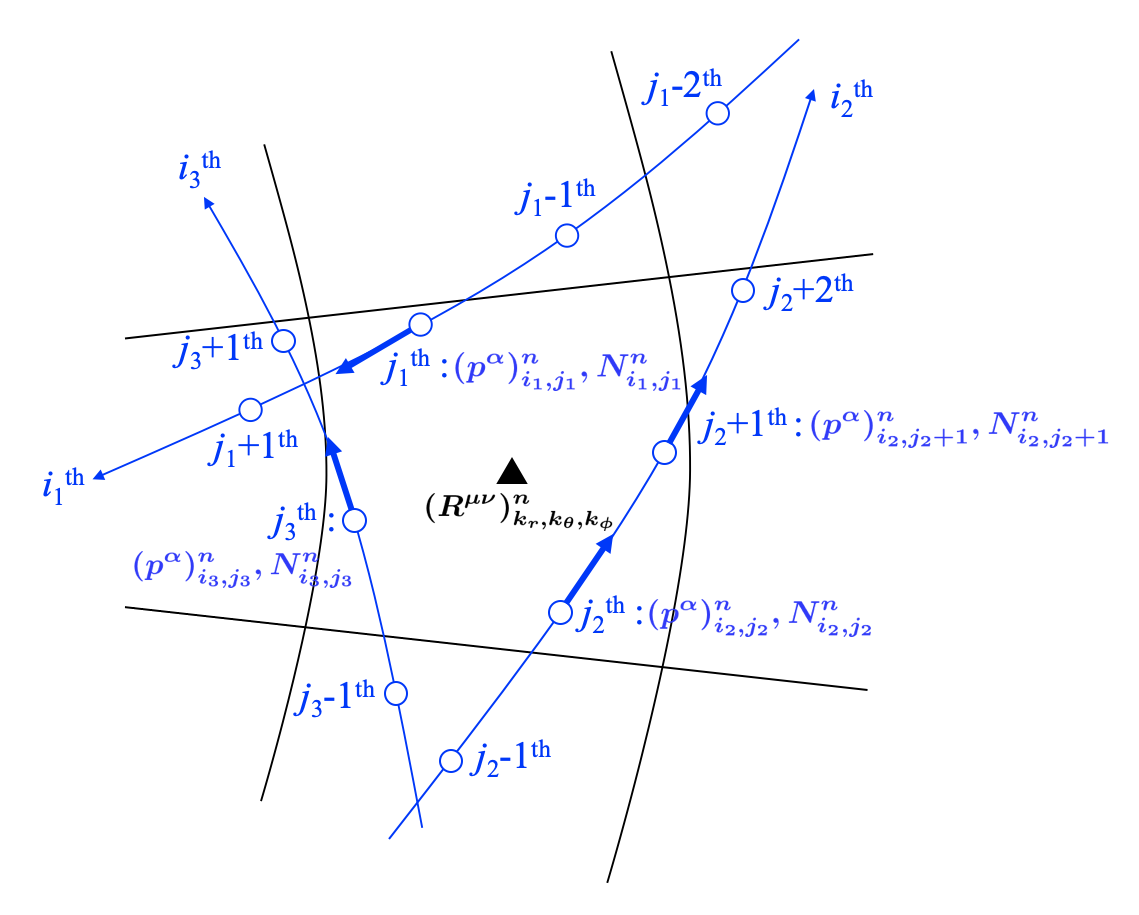}
    \caption{
    Schematic illustration of radiative transfer calculations in \texttt{CARTOON}. Actually, radiative transfer calculations are performed in a 3-dimensional space. However, a 2-dimensional illustration is shown for the explanation in this figure.  
    }
    \label{fig:fig4}
  \end{center}    
\end{figure*}

A schematic illustration of the geodesic grids and the spatial cells is Fig. \ref{fig:fig4}.
Black solid lines show the boundaries of the spatial cells.
Black triangle shows the centre of the spatial cell.
Curved blue arrows show the geodesics, and white circles on the geodesics show the geodesic grids. 
$(p^\alpha)^n_{i,j}$ and $N^n_{i,j}$ denotes the four momentum of the photon and the photon number at the geodesic grid $(i,j)$. Where, the index $n$ denotes the $n$-th time step.

Next, we show a method for evaluating the solid angle defined in the ZAMO frame. 
In Fig. \ref{fig:fig4}, three geodesics pass through the spatial cell $(k_r, k_\theta, k_\phi)$. 
The solid angle of each geodesic in this cell is obtained from the four momentum of the photon at the geodesic grid which has minimum $j$ on the each geodesic, specifically $(i_1,j_1)$, $(i_2,j_2)$, and $(i_3,j_{3})$.
At these three points, we calculate $(\bar{\theta}_{i_1,j_1},\bar{\phi}_{i_1,j_1})$, $(\bar{\theta}_{i_2,j_2},\bar{\phi}_{i_2,j_2})$, $(\bar{\theta}_{i_3,j_{3}},\bar{\phi}_{i_3,j_{3}})$ from the four momentum of the photon in the ZAMO frame $(p^{\hat{\alpha}})^n_{i_1,j_1}$, $(p^{\hat{\alpha}})^n_{i_2,j_2}$, $(p^{\hat{\alpha}})^n_{i_3,j_{3}}$ by Equation (\ref{thbphb}).
Then, Voronoi tessellation is performed with three points as kernels, and an unit sphere is divided into three Voronoi areas,
which correspond to the solid angles in the ZAMO frame.
The solid angles of the $i_1$-, $i_2$-, $i_3$-th geodesics in the spatial cell $(k_r, k_\theta, k_\phi)$ obtained in this way are denoted $\Delta \hat{\Omega}_{i_1,k_r,k_\theta,k_\phi}$,
$\Delta \hat{\Omega}_{i_2,k_r,k_\theta,k_\phi}$,
and $\Delta \hat{\Omega}_{i_3,k_r,k_\theta,k_\phi}$.

If there are multiple geodesic grids on a given geodesic in the spatial cell, the solid angles of those geodesic grids set to be equal. 
For example, the geodesic grids $(i_2,j_2)$ and $(i_2,j_2+1)$ have same solid angle $\Delta \hat{\Omega}_{i_2,k_r,k_\theta,k_\phi}$ (see Fig. \ref{fig:fig4}).
The solid angle in the fluid rest frame obtained from $\Delta \hat{\Omega}_{i,k_r,k_\theta,k_\phi}$ as
\begin{align} \label{dOmega}
  \Delta \tilde{\Omega}_{i,k_r,k_\theta,k_\phi} = \left( \frac{(p^{\hat{t}})^n_{i,j}}{(p^{\tilde{t}})^n_{i,j}} \right)^2 \Delta \hat{\Omega}_{i,k_r,k_\theta,k_\phi}.
\end{align}

\subsection{Radiative transfer}

\label{sec:grrt}
To simplify the situation, we consider a case of pure scattering without emission and absorption in this paper ($\mathcal{E}_{\rm e}=\mathcal{A}_{\rm a}=0$). The scattering coefficient is independent of the frequency ($\sigma_\nu=\sigma$). 
First, we consider isotropic scattering in the fluid rest frame (isotropic scattering in the ZAMO frame is discussed later).
Using Equations (\ref{huhenI}), (\ref{huhenEs}), (\ref{huhenAs}), and (\ref{dOmega}),
Equation (\ref{GRRTeq}) is rewritten with the quantities in the ZAMO frame and in the fluid rest frame as 
\begin{align} \label{mixedGRRTeq}
    \frac{dI_{\hat{\nu}}}{d\hat{s}} 
    = \left( \frac{\hat{\nu}}{\tilde{\nu}} \right)^2 \left[ -\tilde{\rho} \tilde{\sigma} I_{\tilde{\nu}} + \frac{ \tilde{\rho} \tilde{\sigma}}{4\pi} \int I_{\tilde{\nu}} d\tilde{\Omega} \right],
\end{align}
where $\tilde{\rho}$, $\tilde{\sigma}$, $\tilde{\nu}$, $I_{\tilde{\nu}}$, and $d\tilde{\Omega}$ are the gas density, the scattering coefficient, the frequency, the intensity, and the infinitesimal solid angle in the fluid rest frame.
Here, the relation between the separation of the geodesic grids measured in the ZAMO frame $d\hat{s}$ and the affine parameter $d \lambda$ which is $d \lambda = ds / \nu =   d\hat{s}/\hat{\nu} \ (\because ds = dx^t, \nu = p^t)$ is used.

Since the relation between the photon number and the intensity is given by Equations (\ref{huhenI}), (\ref{ItoN}), and (\ref{dV}) as
\begin{align} \label{ItoNApp}
    N = \mathcal{I} \Delta \mathcal{V} = \frac{I_{\hat{\nu}}}{\hat{\nu}^3} \hat{\nu}^2 \Delta \hat{\nu} \Delta\hat{\Omega} \Delta\hat{V} = \frac{I_{\tilde{\nu}}}{\tilde{\nu}^3} \tilde{\nu}^2 \Delta\tilde{\nu} \Delta\tilde{\Omega} \Delta\tilde{V}, 
\end{align}
where $\Delta \mathcal{V}$ is the 6-dimensional phase space volume of the spatial cell.
Equation (\ref{mixedGRRTeq}) is transformed
with multiplying both sides of equation by $\Delta \mathcal{V}$ and using Equation (\ref{ItoNApp}) into
\begin{align} \label{dNds}
    \frac{dN}{d\hat{s}} = \left( \frac{\tilde{\nu}}{\hat{\nu}} \right) \left[ -\tilde{\rho} \tilde{\sigma} N + \frac{\tilde{\rho} \tilde{\sigma}}{4\pi} \int N d\tilde{\Omega} \right].
\end{align}

With ignoring the second term and focusing on the geodesic grid $(i, j)$, we discretize Equation (\ref{dNds}) using $\hat{\nu}=p^{\hat{t}},\tilde{\nu}=p^{\tilde{t}}$ as 
\begin{align} \label{Ndecrease}
    N^{n+1}_{i,j} = N^n_{i,j-1} + &\Delta t \left[ -(\tilde{\rho} \tilde{\sigma} \alpha)^n_{k_r,k_\theta,k_\phi} \left( \frac{(p^{\tilde{t}})^n_{i,j-1}}{(p^{\hat{t}})^n_{i,j-1}} \right) N^n_{i,j-1} \right],
\end{align}
where we use $\Delta \hat{s} = \alpha \Delta s = \alpha \Delta t (\because x^{\hat{t}} = \alpha x^t)$. The interval of the geodesic grids, $\Delta s = c \Delta t = \Delta x^t$, is set to be $0.2M$ as we have mentioned above \footnote{Here, the speed of light $c$ is explicitly written for the explanation.}.
Here, the spacial cell $(k_r, k_\theta, k_\phi)$ includes the geodesic grid $(i,j-1)$.

Equation (\ref{Ndecrease}) implies that while photons travel from geodesic grids $(i,j-1)$ to $(i,j)$, $(\tilde{\rho} \tilde{\sigma} \alpha)^n_{k_r,k_\theta,k_\phi} \left( \frac{(p^{\tilde{t}})^n_{i,j-1}}{(p^{\hat{t}})^n_{i,j-1}} \right) N^n_{i,j-1}$ photons are reduced by scattering.
Therefore, in the spatial cell ($k_r$, $k_\theta$, $k_\phi$), 
\begin{align} \label{NscatApp}
    (N_{\rm scat})^n_{k_r,k_\theta,k_\phi} = \sum_{i^\prime,j^\prime \subset k_r,k_\theta,k_\phi} (\tilde{\rho} \tilde{\sigma} \alpha)^n_{k_r,k_\theta,k_\phi}  \left( \frac{(p^{\tilde{t}})^n_{i^\prime,j^\prime}}{(p^{\hat{t}})^n_{i^\prime,j^\prime}} \right) N^n_{i^\prime,j^\prime}
\end{align}
photons are scattered at each time step.
Here, $\displaystyle \sum_{i^\prime,j^\prime \subset k_r,k_\theta,k_\phi}$ denotes the summation over the geodesic grids in the spatial cell which includes the geodesic grid $(i^\prime,j^\prime)$.
Proportional to the solid angle of the geodesic in the fluid rest frame, we redistribute the scattered photons to all geodesics passing through the spacial cell $(k_r, k_\theta, k_\phi)$.
For geodesics with multiple geodesic grids in the spacial cell, the number of scattered photons is divided equally.
That is, the number of photons increasing by scattering at the geodesic grid $(i,j)$ is given by
\begin{align} \label{Nincrease}
    \frac{\Delta \tilde{\Omega}_{i,k_r,k_\theta,k_\phi}}{4\pi} \frac{1}{J_{i,k_r,k_\theta,k_\phi}} (N_{\rm scat})^n_{k_r,k_\theta,k_\phi},
\end{align}
where 
$J_{i,k_r,k_\theta,k_\phi}$ is the number of the geodesic grids on $i$-th geodesic within the spatial cell ($k_r$, $k_\theta$, $k_\phi$).
For example, in Fig. \ref{fig:fig4}, $J_{i_1,k_r,k_\theta,k_\phi}$
and $J_{i_3,k_r,k_\theta,k_\phi}$ are unity, and we find $J_{i_2,k_r,k_\theta,k_\phi}=2$.
With Equations (\ref{Ndecrease}) and (\ref{Nincrease}), the number of photons at the geodesic grid $(i,j)$ is updated by 
\begin{align} \label{NEQ}
    N^{n+1}_{i,j} = N^n_{i,j-1} + &\Delta t \left[ -(\tilde{\rho} \tilde{\sigma} \alpha)^n_{k_r,k_\theta,k_\phi} \left( \frac{(p^{\tilde{t}})^n_{i,j-1}}{(p^{\hat{t}})^n_{i,j-1}} \right) N^n_{i,j-1} \right. \nonumber \\
    &\left. + \frac{\Delta \tilde{\Omega}_{i,k_r,k_\theta,k_\phi}}{4\pi} \frac{1}{J_{i,k_r,k_\theta,k_\phi}} (N_{\rm scat})^n_{k_r,k_\theta,k_\phi} \right].
\end{align}
An average energy of the scattered photons is obtained by 
\begin{equation} \label{pbar}
    \left< p^{\tilde{t}} \right>^n_{i,j} = \frac{\displaystyle \sum_{i^\prime,j^\prime \subset k_r,k_\theta,k_\phi} \left[ (\tilde{\rho} \tilde{\sigma} \alpha)^n_{k_r,k_\theta,k_\phi} \frac{(p^{\tilde{t}})^n_{i^\prime,j^\prime}}{(p^{\hat{t}})^n_{i^\prime,j^\prime}} (p^{\tilde{t}})^n_{i^\prime,j^\prime} N^n_{i^\prime,j^\prime} \right] }{ (N_{\rm scat})^n_{k_r,k_\theta,k_\phi} }.
\end{equation}
Hence, the time component of the four momentum of the photon at $n+1$-th time step, $(p^{\tilde{t}})^{n+1}_{i,j}$, is calculated as
\begin{align} \label{pnew}
    \left( p^{\tilde{t}} \right)^{n+1}_{i,j} &=
    \frac{1}{N^{n+1}_{i,j}} \left[ \sum_{i^\prime,j^\prime \subset k_r,k_\theta,k_\phi} \left(  (p^{\tilde{t}})^n_{i^\prime,j^\prime} N^*_{i^\prime,j^\prime} \right) \right. \nonumber \\
    &\left. + \frac{\Delta \tilde{\Omega}_{i,k_r,k_\theta,k_\phi}}{4\pi} \frac{1}{J_{i,k_r,k_\theta,k_\phi}} (N_{\rm scat})^n_{k_r,k_\theta,k_\phi} \left< p^{\tilde{t}} \right>^n_{i,j} \right],
\end{align}
because the unscattered photons, of which the average energy is $(p^{\tilde{t}})^n_{i,j}$, and the scattered photons which have the average energy given by Equation (\ref{pbar}) are mixed.
Here, 
\begin{equation} \label{Nstar}
     N^*_{i,j} = N^n_{i,j-1} - \Delta t (\tilde{\rho} \tilde{\sigma} \alpha)^n_{k_r,k_\theta,k_\phi} \left( \frac{(p^{\tilde{t}})^n_{i,j-1}}{(p^{\hat{t}})^n_{i,j-1}} \right) N^n_{i,j-1}
\end{equation}
is the number of the unscattered photons at $n$-th time step.
The space components of the four momentum of the photon, $(p^{\tilde{i}})^n_{i,j}$, are also updated as $(p^{\tilde{i}})^{n+1}_{i,j} = \left[( p^{\tilde{t}} )^{n+1}_{i,j} / (p^{\tilde{t}})^n_{i,j} \right] (p^{\tilde{i}})^n_{i,j}$ to fulfil the normalization condition $p^{\tilde{\alpha}} p_{\tilde{\alpha}}=0$. 

If we consider the isotropic scattering in the ZAMO frame, we replace the physical quantities in the fluid rest frame in Equations (\ref{mixedGRRTeq})-(\ref{Nstar}) with the physical quantities in the ZAMO frame.
For isotropic scattering in other general coordinate systems, the same procedure can be used.

The radiation energy-momentum tensor in the ZAMO frame in the spatial cell ($k_r$, $k_\theta$, $k_\phi$) is obtained using Equations (\ref{huhenI}), (\ref{ItoN}), (\ref{dV}), (\ref{Rmunuhat}), and $p^{\hat{t}}=\hat{\nu}$ as 
\begin{align} \label{defRmunuhat}
    (R^{\hat{\alpha} \hat{\beta}})^n_{k_r,k_\theta,k_\phi} = \sum_{i^\prime,j^\prime \subset k_r,k_\theta,k_\phi} \left( \frac{(p^{\hat{\alpha}})^n_{i^\prime,j^\prime} (p^{\hat{\beta}})^n_{i^\prime,j^\prime} N^n_{i^\prime,j^\prime}}{(p^{\hat{t}})^n_{i^\prime,j^\prime} \Delta V_{k_r,k_\theta,k_\phi}} \right).
\end{align}

{\color{black}
Here we note that the \texttt{CARTOON} algorithm is classified as the ART (Authentic Radiative Transfer) method \citep[][ Nakamoto, Susa \& Umemura in \cite{Iliev2006}]{Razoumov2005}, not the long characteristic method \citep{Abel1999, Sokasian2001, Susa2006, Rijkhorst2006}. 
In the long characteristic method, the radiative transfer equation is solved along geodesics from the boundaries of the computational domain to the centre of all cells. In this way, the intensities at the centre of all cells are calculated. The ART method also solves the radiative transfer equation from the boundaries along the geodesics. However, geodesics do not necessarily have to pass through the centre of cells. Therefore, the intensities at the centre of all cells are evaluated by interpolating from the intensities on the nearest geodesics. The \texttt{CARTOON}'s method is an extension of the ART method that can be used in curved space-time. Although the radiative transfer equation is solved along the curved geodesics, the intensities at the cell centre are calculated by interpolating from the intensities on the nearest geodesics (see Fig. \ref{fig:fig4}).
}

\subsection{Observed intensity map}
Here, we consider the observer's screen of which the centre is located at $(r,\theta,\phi)=(r_{\rm obs}$, $\theta_{\rm obs}$, $\phi_{\rm obs})$. We prepare Cartesian coordinates $(X,Y)$ on the screen,
$-25r_{\rm g}\leq X \leq 25r_{\rm g},\ -25r_{\rm g}\leq Y \leq 25r_{\rm g}$.
Here, the $Y$-axis is parallel to the equatorial plane ($\theta=\pi/2 $ plane). 

We generate the geodesics which are independent of the geodesics generated in Section \ref{sec:geogrid}
by integrating the geodesic equations
backward in time from the observer's screen.
The observer's screen is divided equally into $256\times 256$ and the centres of each pixel are the start points of the integration.
Using the start point $(X,Y)$,
we set initial values of $\bar{\theta},\bar{\phi}$ as
\begin{align} \label{ic_obs}
    \sin \bar{\theta} \sin \bar{\phi} = -\frac{r_{\rm g}}{r_{\rm obs}} X,\ \sin \bar{\theta} \cos \bar{\phi} = \frac{r_{\rm g}}{r_{\rm obs}}Y,
\end{align}
so that each geodesic is perpendicular to the observer's screen.
The initial four momentum of the photon is calculated from Equation (\ref{thbphb}).
In this way, we generate $256\times 256$ geodesics from one observer's screen.

Using Equations (\ref{huhenI}), (\ref{huhenEs}), (\ref{huhenAs}), (\ref{dOmega}) and $d\lambda = d\hat{s}/\hat{\nu} = \alpha dt/\hat{\nu}$, the frequency integrated radiative transfer equation is written as
\begin{equation} \label{GRRTeq_obs}
    \frac{d\hat{I}}{dt} = \left( \frac{p^{\tilde{t}}}{p^{\hat{t}}} \right)^3 \left[ - \tilde{\rho} \tilde{\sigma} \alpha \tilde{I} +
    \frac{\tilde{\rho} \tilde{\sigma}}{4\pi} \alpha R^{\tilde{t} \tilde{t}} \right],
\end{equation}
where, $\hat{I}$ and $\tilde{I}$ are the frequency integrated intensity in the ZAMO frame and the fluid rest frame.
By solving this equation along the geodesics set above, we calculate the intensity map on the observer's screen.

\section{Test calculations}\label{sec:sec3}

We perform the test calculations of the radiation propagation in a 2-dimensional space (equatorial plane)
and compare with the results of TU17 (Section \ref{sec:2d}).
It is shown that the wavefront propagation and the conservation of the number of photons are correctly treated in \texttt{CARTOON}.
Next, we perform the test calculations of the radiation propagation in a 3-dimensional space (Section \ref{sec:3d}).
In addition to the wavefront propagation, we also show the time variation of the intensity map on the observer's screen (Section \ref{sec:obs}).
In this section, the gas densities in the ZAMO frame $\hat{\rho}$ and in the fluid rest frame $\tilde{\rho}$ are assumed to be uniform and time-invariant.

\subsection{Two-dimensional radiation propagation test}
\label{sec:2d}
Although \texttt{CARTOON} is a 3-dimensional code, here we perform 2-dimensional test calculations for comparison with TU17.
For this purpose, we solve the radiation transfer on the equatorial plane ($\theta=\pi/2$).
The number of the spatial cells is $N_r \times N_\phi = 120 \times 128$ and 
we set $N^{\rm geo}_{\phi}=128$.
We generate geodesics similar to TU17.
The BH spin parameter is assumed to be $a=0.5$.
Initially a uniform and isotropic radiation field is set up within a region of radius $1.0r_{\rm g}$ centred at $(r,\phi)=(6.0 r_{\rm g},0.0)$.

\begin{figure}
	\includegraphics[width=\linewidth]{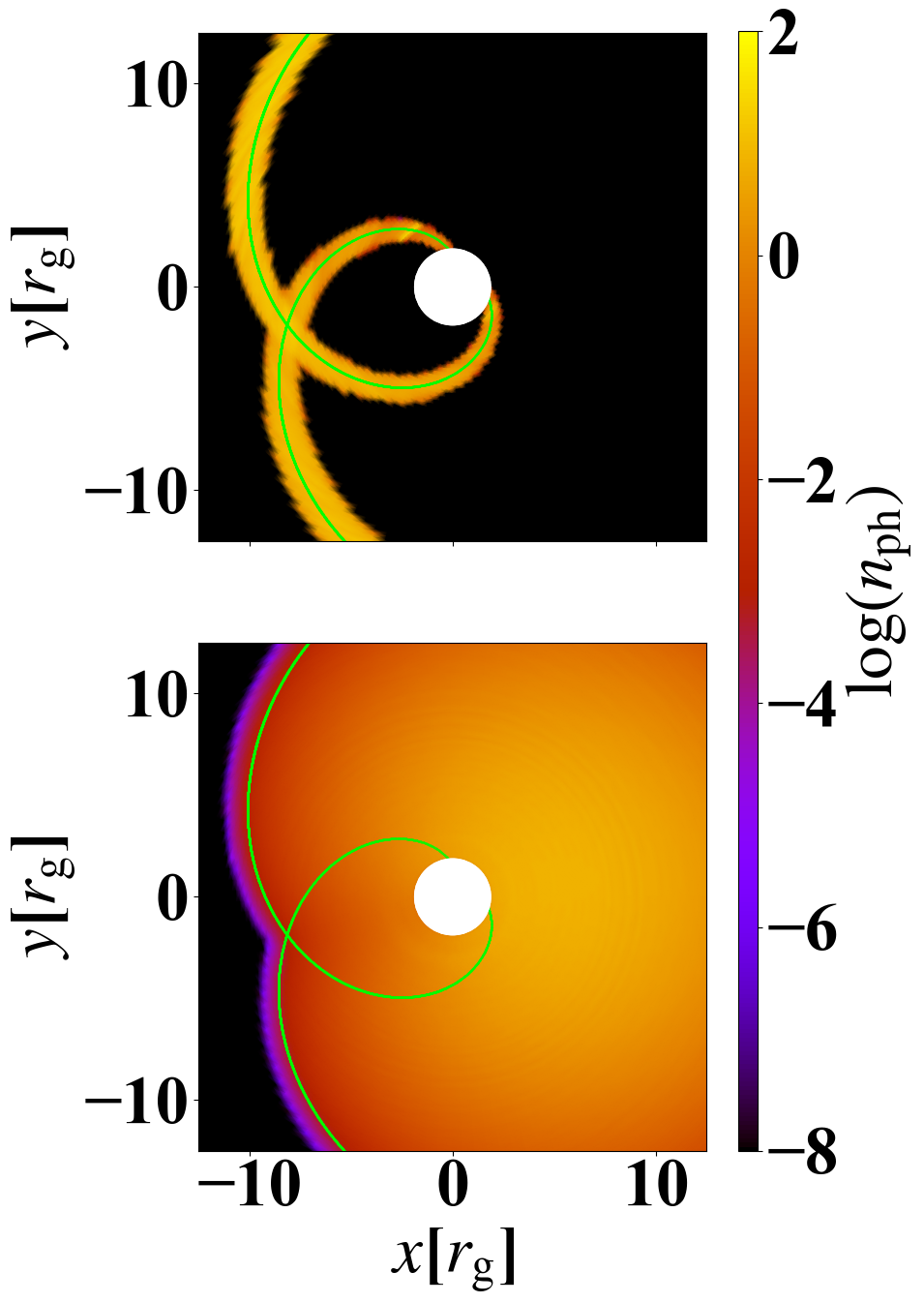}
    \caption{
    The photon number density, $n_{\rm ph}$, on $x$-$y$ plane (equatorial plane) at $t=24t_{\rm g}$ for the 2-dimensional wavefront propagation test, where $x=r\cos\phi$ and $y=r\sin\phi$.
    The top and bottom panels show the results without scattering and with isotropic scattering in the ZAMO frame. 
    Green solid lines denote the wavefronts at $t=24t_{\rm g}$ associated with the isotropic emission from the point source at $(r,\phi)=(6.0 r_{\rm g},0.0)$. 
    White circles denote the BH.
    }
    \label{fig:fig5}
\end{figure}
Fig. \ref{fig:fig5} shows the distribution of the photon number density (color contour)  
and the wavefront associated with the isotropic emission from the point source at the centre of the initial radiation field \citep[see also ][]{Hanni1977, Takahashi1990}
at $t=24t_{\rm g}$. 
The radiation does not propagate isotropically because of the light bending. 
A part of the wavefront rotates clockwise around the BH and another part rotates counter-clockwise.
Thus, the wavefronts intersect each other in the left side (where $x<0$) of Fig. \ref{fig:fig5}.
The part of the wavefront closer to the BH than the intersection (the point where the two green lines cross) has already crossed over.

The top panel of Fig. \ref{fig:fig5} presents the result without scattering. 
We can see the wavefront is in good agreement with the green solid line, 
meaning that \texttt{CARTOON} accurately calculates the propagation of the wavefront.
{\color{black}
}
Note that in this test, some photons are ahead of the green line. This is because the radiation is initially placed in the circular region, so there are photons that travel ahead of the wavefront indicated by the green line by the radius of the circle.
We also stress that the wavefront intersections are correctly reproduced, although the crossing of the radiation cannot correctly solved by the moment method (e.g. M1 closure method).
In the bottom panel of Fig. \ref{fig:fig5}, we employ the isotropic scattering in the ZAMO frame with the scattering coefficient $\hat{\sigma}=0.1$.
The position of the wavefront is the same as the bottom panel. However, the photon number density at the wavefront is reduced via the scattering and the scattered photons are distributed inside the wavefront (the region through which the wavefront has already passed). 

\begin{figure}
	\includegraphics[width=\linewidth]{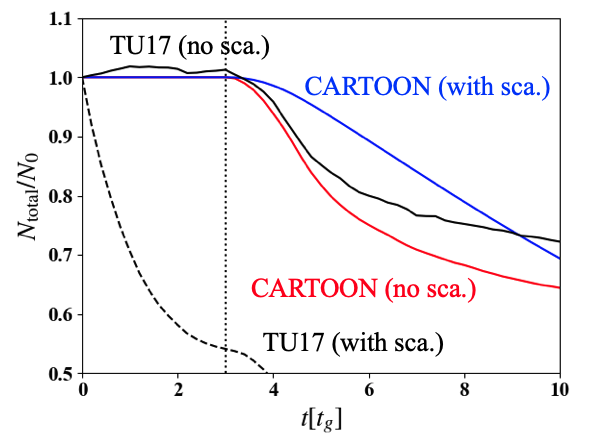}
    \caption{
    Time evolution of the photon number in the computational domain ($N_{\rm total}$) normalized by the initial photon number ($N_0$) in the 2-dimensional radiation propagation test. The red and blue lines show the results of \texttt{CARTOON} without and with scattering. The solid black line and the dashed black line show the result of TU17 without and with scattering. The vertical dotted black line represents the time the wavefront reaches the event horizon.
    }
    \label{fig:fig6}
\end{figure}

Although Fig. \ref{fig:fig5} seems to be very similar to Fig. 7 of TU17, the number of photons obtained in \texttt{CARTOON} is different from that in TU17.
Fig. \ref{fig:fig6} represents a ratio of the photon number in the computational domain,
\begin{align} \label{Ntotal}
    N_{\rm total} = \sum_{i, j} N^n_{i,j},
\end{align}
to the initial photon number $N_0$.
Fig. \ref{fig:fig6} shows that both the red and blue lines remain 1.0 until they intersect the vertical black dotted line.
This means that the number of photons does not change before the photons are swallowed by the BH, since the vertical dotted line indicates the time at which the wavefront reaches the event horizon. 
The decrease in the number of photons at $t \gtrsim 3t_{\rm g}$ occurs because photons enter the event horizon
 (note that the wavefront does not reach the outer boundary until $t \sim 25t_{\rm g}$).
In the case without scattering, photons tend to concentrate around the wavefront.
For this reason, the number of photons decreases more quickly in the case without scattering than in the case with scattering.
In contrast, TU17 does not accurately treat the photon number conservation (see black solid line and black dashed line).
This is clearly understood from the fact that $N_{\rm total}/N_0$ is not retained at $t \lesssim 3t_{\rm g}$.
We stress again that \texttt{CARTOON} guarantees photon number conservation and accurately solves the radiation propagation.

\subsection{Three-dimensional radiation propagation test}
\label{sec:3d}
In this test, the number of the spatial cells is $ N_r \times N_\theta \times N_\phi = 60 \times 32 \times 64$. The BH spin parameter $a$ is assumed to be $a=0.5$. 
We initially set a uniform and isotropic radiation field in the region with radius $1.0r_{\rm g}$ centred at $(r,\theta,\phi)=(6.0 r_{\rm g}, \pi/2, 0)$.
Here, we consider three cases that are without scattering, with the isotropic scattering in the ZAMO frame, and the isotropic scattering in the fluid rest frame.
The scattering coefficient is $\hat{\sigma}=\tilde{\sigma}=0.1$.
When we consider the isotropic scattering in the fluid rest frame, we assume that the fluid motion is the circular motion around the rotation axis of the BH ($\theta=0$ and $\pi$).
The rotation velocity is set to depend only on the distance from the rotational axis of the BH, $r_{\rm cyl}=r\sin\theta$.
At the region of $r_{\rm cyl}\geq r_{\rm ISCO} (\simeq 4.2 r_{\rm g} {\rm \ for\ } a=0.5)$, with $r_{\rm ISCO}$ being the radius of the innermost stable circular orbit (ISCO), we adopt the Keplerian velocity at the equatorial plane regardless of altitude.
On the other hand, the velocity at the region of $r_{\rm cyl}<r_{\rm ISCO}$ is set to be the Keplerian velocity at the ISCO.
This velocity profile is partly modified from the Keplerian shell model \ \citep{Cunningham1975,Falcke2000,Broderick2005-10}.
Here we note that more geodesics must be provided to cleanly trace the wavefront for the case without scattering
since the region where the photon is initially set is sufficiently small relative to the size of the computational domain.
However, setting up an extremely large number of geodesics makes numerical calculations significantly more difficult.
Therefore, we prepared only the geodesics needed to perform this test.
We adopt almost 15 million geodesics, starting from the inside of the sphere where the radiation is initially set.

\begin{figure*}
	\includegraphics[width=\linewidth]{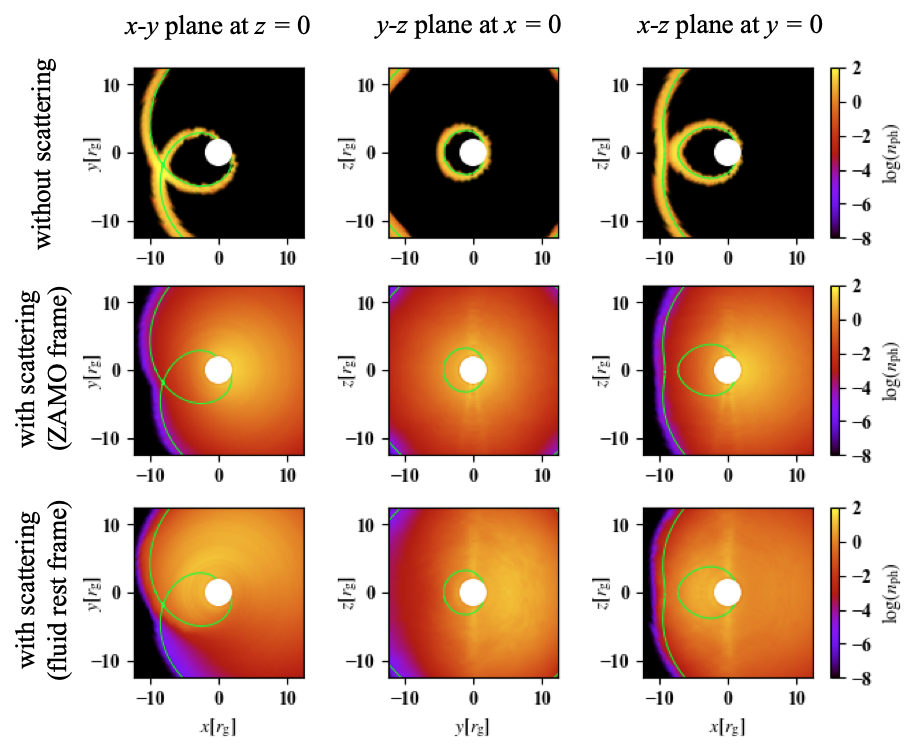}
    \caption{
    The photon number density on $x$-$y$ plane at $z=0$ (left), $x$-$z$ plane at $y=0$ (middle), and $y$-$z$ plane at $x=0$ (right) at $t=24t_{\rm g}$ in the 3-dimensional radiation transfer test, where $x=r\sin\theta\cos\phi$, $y=r\sin\theta\sin\phi$, and $z=r\cos\theta$. The top, middle, and bottom panels show the results without scattering, with isotropic scattering in the ZAMO frame, and with isotropic scattering in the fluid rest frame respectively. 
    Green solid lines denote the wavefronts at $t=24t_{\rm g}$ associated with the isotropic emission from the point source at $(r,\theta,\phi)=(6.0 r_{\rm g}, \pi/2, 0)$. {\color{black} White circles denote the BH.}
    }
    \label{fig:fig7}
\end{figure*}
Fig. \ref{fig:fig7} shows the distribution of the photon number density (color contour) and the wavefront of isotropic radiation from the point source at the centre of the initial radiation field at $t=24t_{\rm g}$.
In the case without scattering, photons concentrate around the wavefront (see top panels).
In addition, there are no photons in the regions where wavefront has not yet reached (the reason for the slight deviation from the green line has been mentioned in Section \ref{sec:2d}).
These imply that \texttt{CARTOON} successfully reproduces the wavefront propagation in 3-dimensional space.
We note that the distribution of the photon number density without scattering, which is a little uneven, becomes smooth according as the number of the geodesics increases.

In the cases with the scattering, it is found that the scattered photons fill up the region where the wavefront has passed. 
In the calculation with the isotropic scattering in the ZAMO frame, the high photon number density region (yellow region) in the $x$-$y$ plane at $z=0$ is located in the right of the BH. 
On the other hands, the calculation with the isotropic scattering in the fluid rest frame, that region is located in the top left to the top of the BH.
This is because the fluid rotate counter-clockwise in the $x$-$y$ plane at $z=0$ and because the isotropic scattering in the fluid rest frame works to advect the photons towards the direction of the fluid motion.
{\color{black}
The high number density regions along the $z$-axis in the $y-z$ plane and $x-z$ plane of Fig. \ref{fig:fig7} are numerical artifacts (see Appendix \ref{sec:AppB}).
}

\begin{figure}
	\includegraphics[width=1.05 \linewidth]{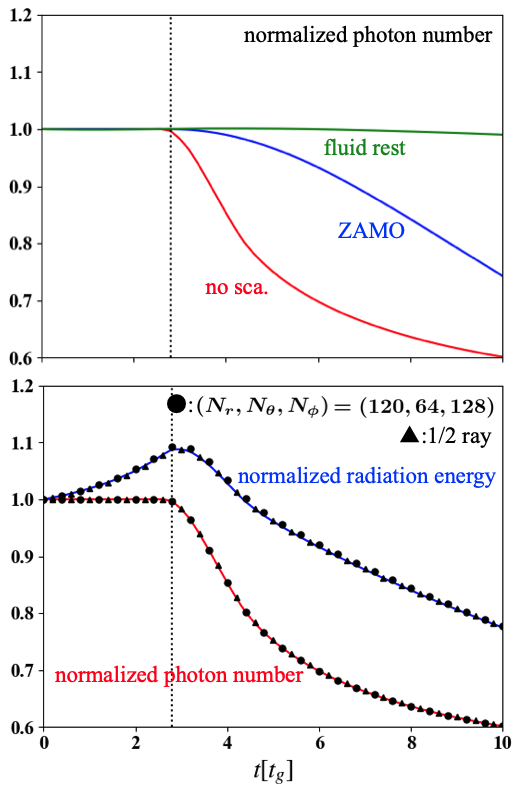}
    \caption
    {
    The top panel shows the time evolution of the photon number in the computational domain normalized by the initial photon number in the 3-dimensional radiation propagation test. The red, blue, and green lines represent the results without scattering, with a scattering in the ZAMO frame, and a scattering in the fluid rest frame. The wavefront reaches the event horizon at around $t\sim 3t_{\rm g}$ (dotted black line).
    The lower red line on the bottom panel is the same as the red line on the top panel.
    The bottom panel also shows the radiation energy in the computational domain normalized by the initial radiation energy without scattering (upper blue line). 
    Black circles and triangles represent the results of increasing the number of the spatial cells ($(N_r,N_\theta,N_\phi)=(120,64,128)$) and reducing the number of the geodesics by about half, respectively.
    }
    \label{fig:fig8}
\end{figure}
The top panel of Fig. \ref{fig:fig8} shows the time evolution of the photon number in the computational domain normalized by the number of photons initially given.
Until the wavefront reaches the event horizon at $t\sim 3t_{\rm g}$ (vertical dotted black line), the photon number keeps constant in all cases as in the 2-dimensional test calculations (see Section \ref{sec:2d}). 
We can see that the number of photons rapidly decreases just after the wavefront reaches the event horizon in the case without scattering ($t\gtrsim 3t_{\rm g}$) since photons concentrate around the wavefront since photons concentrated around the wavefront rapidly fall onto the BH.
This trend is also similar to the 2-dimensional test calculation. 
We also find that photon number decreases very slowly in the calculation with isotropic scattering in the fluid rest frame.
This is because in the case of isotropic scattering in the fluid rest frame, the number of photons going to the event horizon is much smaller than in the case of the isotropic scattering in the ZAMO frame.
Photons tend to be transported in the $\phi$-direction when isotropically scattered by a rotating fluid.

The conservation of the number of photons is guaranteed by \texttt{CARTOON} even if the number of geodesics is reduced.
In the bottom panel, we plot the photon number when the number of geodesics is reduced by half (lower triangles).
We find that the number of photons varies in the same way regardless of the number of geodesics.
The upper blue line represents the radiation energy in the computational domain normalized by the initial radiation energy
for the case without scattering.
At $t \lesssim 3t_{\rm g}$, the radiation energy increases because many photons approach the event horizon.
However, the radiation energy decreases at $t \gtrsim 3t_{\rm g}$ because photons are swallowed by the BH.
The behaviour of the radiation energy does not change so much even if the number of geodesics is reduced.
It is clearly understood by that the upper blue line and the upper triangles almost overlap.
This panel also shows that the number of the spatial cells is sufficient at least in this test.
The results of increasing the number of the spatial cells by a factor of 8 (black circles) is consistent with the original results (red and blue lines).

\subsection{Observed images}
\label{sec:obs}

Finally, we show the intensity map on the observer's screen for the case of isotropic scattering in the fluid rest frame in the 3-dimensional radiation propagation test (Section \ref{sec:3d}).

\begin{figure*}
	\includegraphics[width=\linewidth]{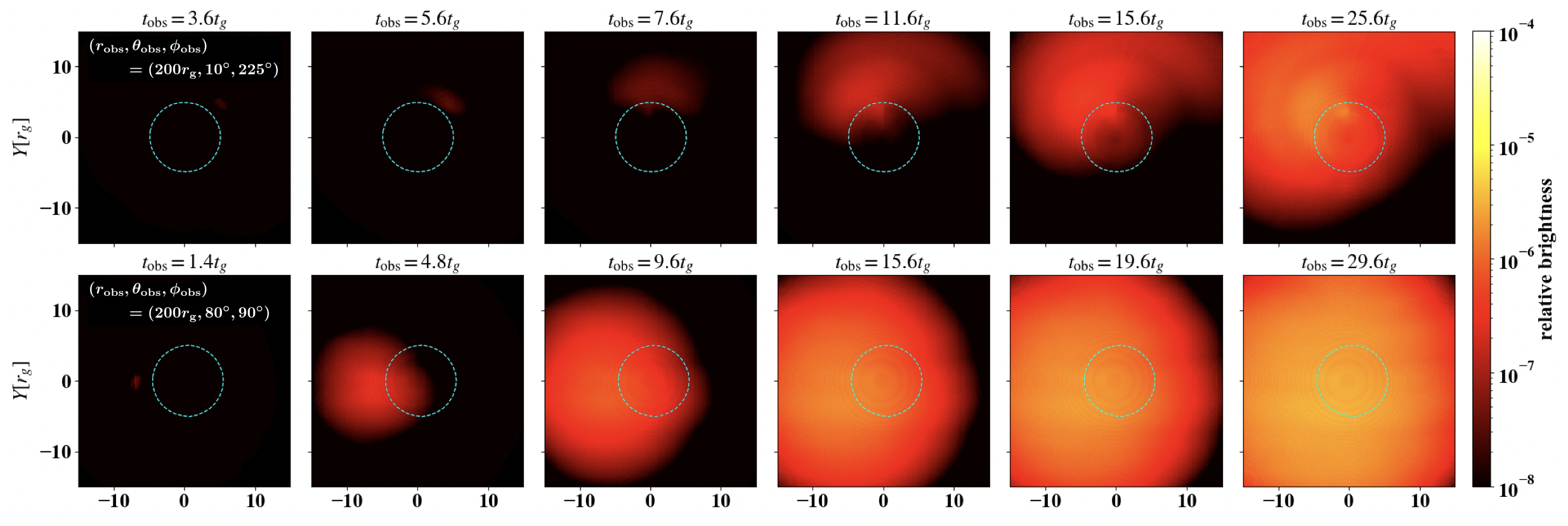}
    \caption{
    Time variations of the intensity map on the observer's screen at $(r_{\rm obs},\theta_{\rm obs}, \phi_{\rm obs})=(200r_{\rm g}, 10^\circ,225^\circ)$ (top raw) and $(200r_{\rm g}, 80^\circ,90^\circ)$ (bottom raw).
    Cyan dashed circles denote the BH shadow analytically obtained \citep{Bardeen1973, Chandrasekhar1983, Takahashi2004}.
    Here, $t_{\rm obs}$ is the time measured from the moment the first photon arrives at point of $(r_{\rm obs}, \theta_{\rm obs}, \phi_{\rm obs})=(200r_{\rm g}, 80^\circ,90^\circ)$ ($\tilde{\sigma}=0.1$).
    }
    \label{fig:fig9}
\end{figure*}
The top raw of Fig. \ref{fig:fig9} is the intensity map on the screen at $(r_{\rm obs}, \theta_{\rm obs}, \phi_{\rm obs})=(200r_{\rm g}, 10^\circ,225^\circ)$ (almost face-on view).
We can see that the bright region expands and rotates counter-clockwise since the photons tend to be transported in the $\phi$-direction with fluid as we have mentioned above.
On the other hand, the left side of the screen becomes brighter at first and the bright area gradually expands with time
on the screen of $(r_{\rm obs}, \theta_{\rm obs}, \phi_{\rm obs})=(200r_{\rm g}, 80^\circ,90^\circ)$ (almost edge-on view).

\begin{figure}
  \begin{center}
	\includegraphics[width=\columnwidth]{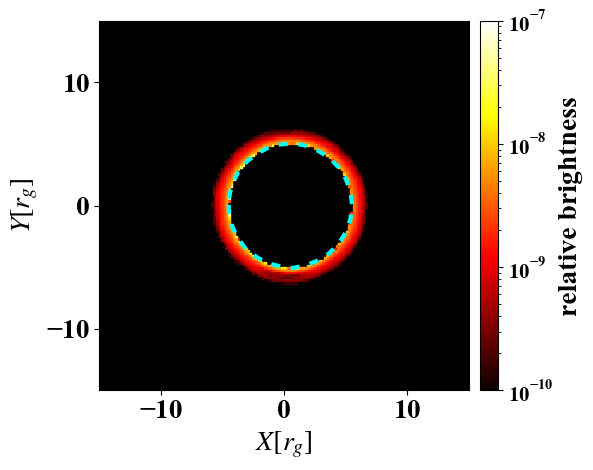}
    \caption{
    The intensity map on the observer's screen at $(r_{\rm obs}, \theta_{\rm obs}, \phi_{\rm obs})=(200r_{\rm g}, 80^\circ,90^\circ)$ and $t_{\rm obs} = 56t_{\rm g}$ for $\tilde \sigma=0.01$. Cyan dashed circle denotes the BH shadow same as Fig. \ref{fig:fig9}.
    }
    \label{fig:fig10}
  \end{center}
\end{figure}

Here, we note that the photon ring does not appear in Fig. \ref{fig:fig9} because of the large optical thickness (scattering coefficient).
In this case, the last scattering surface is located between the screen and the BH. 
Thus, many scattered photons reach the observer's screen.
The intensity enhanced by the unstable circular orbit around the BH is attenuated and no clear shadow appears on the screen.
However, the BH shadow is visible by reducing the scattering coefficient.
This is clearly shown in Fig. \ref{fig:fig10}.
{\color{black}
Here, $\sigma$ is reduced to 0.01. The radiation fields are fixed at the result at $t=100 t_{\rm g}$ in the test calculation shown in Section \ref{sec:3d} for $20t_{\rm g}$ and then set to zero. With this setting, until $t_{\rm obs} \lesssim 30 t_{\rm g}$, a region of high intensity, exceeding $\sim 10^{-7}$ , appears in the area of radius $5r_{\rm g}$ around the centre of the observer’s screen. The photon ring (BH shadow) is not visible. Thereafter, the intensity gradually weakens and the photon ring clearly appears at around $t_{\rm obs}=56t_{\rm g}$. Fig. \ref{fig:fig10} shows the intensity map at this time.
}

\section{Conclusions \& Discussion}
\label{sec:sec4}
In the present study, we have developed the 3-dimensional general relativistic radiation transfer (GRRT) code: \texttt{CARTOON} by significantly improving the 2-dimensional GRRT code: \texttt{ARTIST} (TU17).
In \texttt{CARTOON}, the radiative transfer equation is solved along the geodesics in the photon number conservation scheme. 
This allows us not only to calculate the propagation of the wavefront accurately {\color{black} but also to guarantee the conservation of photon number}.
Here, the geodesics reaching equally spaced inner and outer boundaries are employed.
The Voronoi tessellation method is used to set geodesics and evaluate solid angles.
Also, we implement the isotropic and coherent scattering in the ZAMO frame and in the fluid rest frame.
The change of the radiation energy due to the scattering is also treated correctly.

We have conducted the test calculations of radiation propagation in a 2-dimensional and 3-dimensional space. 
We have confirmed that the propagation of the wavefront can be solved correctly in both cases with and without scattering (correspond to the case in a vacuum) in \texttt{CARTOON}.
Our code can also properly calculate the number of photons and the intersection of the wavefront that behave unphysically in the M1 closure method.
In addition, we successfully have demonstrated the time variation of the intensity map on the observer's screen simultaneously with the evolution of the radiation field around the BH.

Although  we have not yet implemented the emission and the absorption in the present code,
these implementations are straightforward.
The change in photon number, which is estimated from the difference between emission and absorption, should be solved for together with the decrease in the number of photons due to scattering.
Since the absorption coefficient is given in the fluid rest frame, 
it would be better to use the radiative transfer equation which the right-hand side is written in the fluid rest frame.
It is left as important future work.
In the present study, we have solved the frequency-integrated radiative transfer equation. 
In order to obtain the frequency-dependent intensity,
we should solve the frequency-dependent radiative transfer equation in the photon number conservation scheme for each frequency.
The calculation of average energy of the scattering photons (Equation (\ref{pbar})) is no longer needed. 
If \texttt{CARTOON} is updated into a frequency-dependent code with emission/absorption via the synchrotron as well as bremsstrahlung and Compton scattering, it would be possible to calculate the spectra and the observed images at each wavelength of the accretion discs around the BHs.

{\color{black}
Also, the way geodesics are set up may need to be improved. The geodesics should be generated as even as possible in the whole of the computational domain and a sufficient number of geodesics pass through each cell. In order to reduce the computational cost, the number of geodesics should be small. As an example that fulfils these conditions relatively well, we generate geodesics by the method described in Section \ref{sec:geogrid}. In the present method, almost 60\% of geodesics generated according to the method described in Section \ref{sec:geogrid} orbit less than one time, almost 20\% of geodesics orbit between one and two times and almost 20\% of geodesics orbit between two and three times around the BH. In the test calculation described in the later Section \ref{sec:sec3}, our results are good enough with geodesics generated by the method described in Section \ref{sec:geogrid}. However, increasing the number of geodesics or using another method may be appropriate depending on the problem. We think studying the best way to generate geodesics is an important future issue.
}

{\color{black} Finally, we have solved the GRRT equation} by assuming the steady flow in this work.
However, by coupling \texttt{CARTOON} with magnetohydrodynamics code, GRRMHD simulations can be performed.
Variable Eddington Tensor method is one of the ways. 
In \citet{Asahina2020}, the Eddington tensor is computed by solving the grid-based GRRT equation (see also \citet{Asahina2022}).
If we solve the radiative transfer equation by \texttt{CARTOON} and determine the Eddington tensor, the accuracy of GRRMHD calculations will be improved.
It is also possible to carry out the simulations in which the radiation moment is directly evaluated from the intensity without the moment equation. Such simulations have already been performed, although general relativity is not incorporated in \citet{Ohsuga2016, Jiang2014a, Jiang2014b, Jiang2014c}.
{\color{black}
The geodesic grids are set with equal intervals ($ds = 0.2M$) and the time step is constant at $0.2M$ in the present paper. Then, we adopt a perfect up-wind method. We can consider two methods when we perform the radiation hydrodynamics simulation by coupling \texttt{CARTOON} with hydrodynamics code. One is to use the time step determined by the interval of geodesic grids. The other is to use the time step determined by the CFL condition of the hydrodynamics simulation. If we use the former method, the interval of geodesic grids should be small to avoid breaking the CFL condition. On the other hand, if we use the latter method, we need to improve CARTOON to solve the radiative transfer equation with the variable time step. Also, we might need to adjust how to generate geodesic grids. We think that these are important future works.
}

\section*{Acknowledgements}
We would like to thank the anonymous reviewer for invaluable comments
and useful suggestions. We thank Kohji Yoshikawa for providing \texttt{Icosahedron-based Geodesic Dome}.
Our simulations were conducted with PPX at the Center for Computational Sciences (CCS), University of Tsukuba.
This work was supported by JSPS KAKENHI Grant numbers 22J10256 (M.M.T.), 18K03710, 21H04488 (K.O.), 19H00697 (M.U. and K.O.), 21H01132 (R.T., M.U., K.O. and Y.A.), 18K13581 (Y.A.) and Multidisciplinary Cooperative Research Program in CCS, University of Tsukuba.
This work was also supported (in part) by MEXT as ”Program for Promoting Researches on the Supercomputer Fugaku” (Toward a unified view of the universe: from large scale structures to planets, JPMXP1020200109) (K.O.), and by Joint Institute for Computational Fundamental Science (JICFuS, K.O.), Multidisciplinary Cooperative Research Program in CCS, University of Tsukuba.

\section*{Data Availability}
The data underlying this article will be shared on reasonable request to the corresponding author.



\bibliographystyle{mnras}
\bibliography{cartoon} 



\appendix
{\color{black}
\section{Setting of geodesics and accuracy}
\label{sec:AppA}
In the paper, almost 20\% of the geodesics generated by the method described in section \ref{sec:geogrid} orbit the BH twice, and another 20\% orbit the BH once. Geodesics that can orbit the BH more than three times are not included. We investigate how long can calculate the time evolution of the radiation energy correctly in the wavefront propagation test in the setting of this study (see Section \ref{sec:3d}). 

\begin{figure}
  \begin{center}
	\includegraphics[width=\columnwidth]{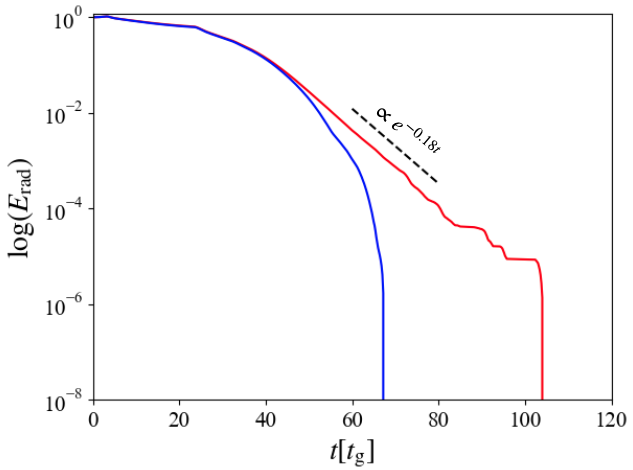}
    \caption{
    Time evolution of the radiation energy in the computational domain in the 3-dimensional wavefront propagation shown in Section \ref{sec:3d}. The red line is the result with the present method of not using geodesics that orbit the BH more than three times. The blue line shows the result obtained without geodesics that orbit the BH more than twice. Theoretically, the radiation energy in the computational domain decreases in proportion to $e^{-0.18t}$ (black dotted line).
    }
    \label{fig:figapp1}
  \end{center}
\end{figure}

Fig. \ref{fig:figapp1} shows the result of this. In this test, it is known that the radiation energy in the computational domain decreases exponentially with time (see also the black dotted line in Fig. 8 in TU17 for the case of BH spin $a=0.5$). We can calculate the time evolution correctly from the beginning of the calculation to about $100t_{\rm g}$ with geodesics set by the method described in the paper. However, the radiation energy rapidly decreases after $100t_{\rm g}$. This means that approximately all photons have reached the event horizon or escaped from the system by $t \sim 100t_{\rm g}$, since no geodesic has been set for more than three orbits around the BH. Furthermore, when we exclude geodesics that orbit more than twice around the BH, we can correctly calculate the propagation of the wavefront for only about $40t_{\rm g}$. This is because photons can only circle the BH twice. Thus, in order to solve the propagation of photons more accurately, we need to use more geodesics that are close to the photon ring and orbit the BH many times.

\section{Numerical artifacts along the z-axis}
\label{sec:AppB}
The high photon number density regions along the $z$-axis in the $y-z$ plane and the $x-z$ plane of Fig. \ref{fig:fig7} are numerical artifacts. The reason that the artifacts appear is that the ratio of the number of geodesic grids in the spatial cell on the z-axis to that surrounding it, $\mathcal{R}$, is very large. Therefore, they disappear by adjusting the number of geodesics and spatial cells and making $\mathcal{R}$ smaller. This is shown by the following tests.

\begin{figure}
  \begin{center}
	\includegraphics[width=\columnwidth]{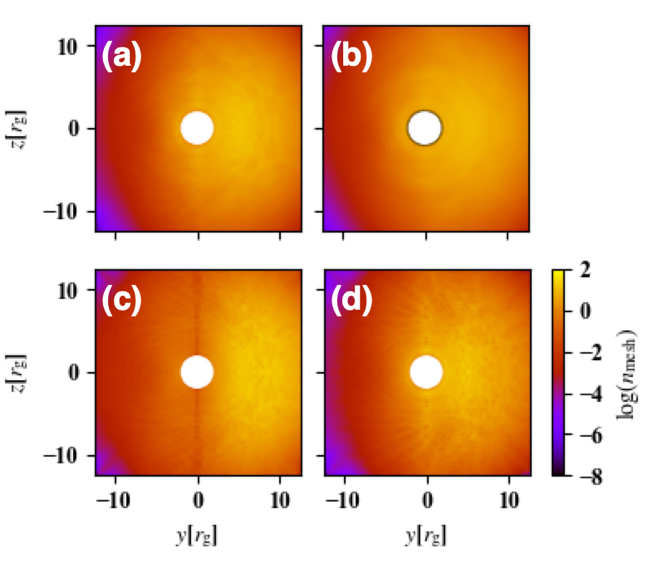}
    \caption{
    Same as bottom centre panel of Fig. \ref{fig:fig7}, but (a) with half the number of geodesics, (b) with fewer spatial cells ($N_r \times N_\theta \times N_\phi = 30\times16\times32$), (c) with more spatial cells ($N_r \times N_\theta \times N_\phi = 120\times64\times128$) and (d) with more spatial cells ($N_r \times N_\theta \times N_\phi = 120\times64\times128$) and half the number of geodesics.
    }
    \label{fig:figapp2}
  \end{center}
\end{figure}
Fig. \ref{fig:figapp2} (a) shows the result of the test calculation same as Section \ref{sec:3d}, but with half the number of geodesics. In this case, there is no numerical artifact. This is because $\mathcal{R}$ in Fig. \ref{fig:figapp2} (a) is smaller ($\mathcal{R}=3.02$) than Fig. \ref{fig:fig7} ($\mathcal{R}=3.25$). In addition, Fig. \ref{fig:figapp2} (b) and (c) show the results with fewer spatial cells ($N_r \times N_\theta \times N_\phi = 30\times16\times32$) and more spatial cells ($N_r \times N_\theta \times N_\phi = 120\times64\times128$). Although there is no numerical artifact in the former ($\mathcal{R}=2.06$), a numerical artifact appears in the latter case ($\mathcal{R}=3.34$), and the photon number density becomes lower along the $z$-axis. Fig. \ref{fig:figapp2} (d) shows the result with more spatial cells ($N_r \times N_\theta \times N_\phi = 120\times64\times128$) and half the number of geodesics. In this case, there are no numerical artifacts ($\mathcal{R}=2.76$). Therefore, we conclude that the ratio of the number of geodesic grids in the spatial cell surrounding (tangent to) the $z$-axis and on the $z$-axis, $\mathcal{R}$, should be reduced by adjusting the number of geodesics and spatial cells. Table \ref{tab:tabapp2} is the summary of the results of Appendix. \ref{sec:AppB}. 

\begin{table}
 \caption{Summary of the results of Appendix. \ref{sec:AppB}}
 \label{tab:tabapp2}
 \begin{tabular}{cccc}
  \hline
  $N_r \times N_\theta \times N_\phi$ & Number of geodesics & $\mathcal{R}$ & Numerical artifact\\
  \hline
  $60\times32\times64$ & fidutial & 3.25 & YES\\
  $60\times32\times64$ & half & 3.02 & NO\\
  $30\times16\times32$ & fidutial & 2.06 & NO\\
  $120\times64\times128$ & fidutial & 3.34 & YES\\
  $120\times64\times128$ & half & 2.76 & NO\\
  \hline
 \end{tabular}
\end{table}

In addition, we investigate the effect of the BH spin and find that the BH spin is not the direct cause of numerical artifacts. It is obvious from the appearance of numerical artifacts in Fig. \ref{fig:figapp3} (a). 

\begin{figure}
  \begin{center}
	\includegraphics[width=\columnwidth]{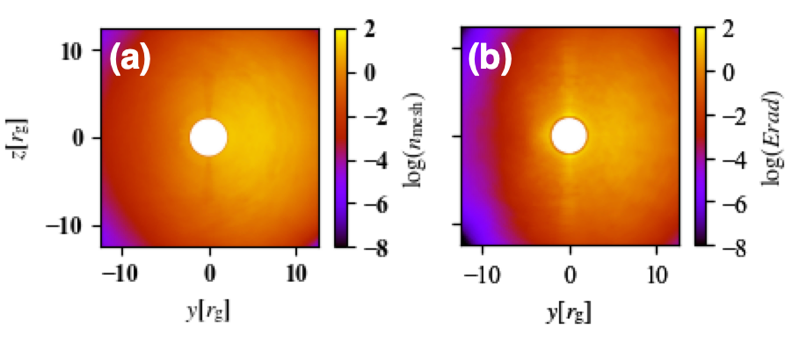}
    \caption{
    (a) Same as bottom centre panel of Fig. \ref{fig:fig7}, but the BH spin parameter is $a = 0.0$, (b) Radiation energy density ($R^{tt}$) obtained from the photon number density shown in the lower centre panel of Fig. \ref{fig:fig7}.
    }
    \label{fig:figapp3}
  \end{center}
\end{figure}

Fig. \ref{fig:figapp3} (a) shows the result same as Section \ref{sec:3d}, but with the BH spin $a=0.0$. Clearly, a numerical artifact appears. We also plot the distribution of the radiation energy density with the BH spin $a=0.5$ in Fig. \ref{fig:figapp3} (b). It is found that a numerical artifact appears as with the distribution of the photon number density.
}

\bsp	
\label{lastpage}
\end{CJK}
\end{document}